# Influence of Ionic strength on calcium carbonate (CaCO$_3$) polymorphism

**Taylor Evans\***
**\***Department of Chemistry, University of Michigan, 930 N. University Ave., Ann Arbor, MI 48109
*E-mail: taevans@umich.edu (T.E.)

## ABSTRACT

CaCO$_3$ crystals' physical properties, such as polymorphism and hence the reflectivity and stability, are critical factors of their qualities in industrial applications. Factors such as additives and substrates that influence CaCO$_3$ polymorphism have been intensively studied (Becker et al., 2005; Kim et al., 2005). However, the effects of ionic strength created by varying additives are seldom paid attention to. This study is analyzing how ionic strength of the growth solution influences the crystalline structure of CaCO$_3$, by applying growth solutions containing different types of cations of varying concentrations, K$^+$, Na$^+$, and NH$_4^+$. This study reveals that the ionic strength plays a significant role in polymorph selection in the way that the percentage of vaterite among the precipitates increases with the concentration of ionic strength.

## INTRODUCTION

The investigations of biomineralization are relatively new to the scientific world and the mechanisms behind it are still under intense research for the full understanding of how the growth environment influences the biominerals' development (Addadi and Weiner, 1985; Hu et al., 2010; Mann, 2001). One such group of biominerals are of particular interest, calcium carbonates (CaCO$_3$) found in the hard exoskeletons of marine organisms. The impact of ion additives in the growth environment, such as Mg$^{2+}$ (Davis et al., 2000) on CaCO$_3$ morphology have drawn a high attention. However, the effects of ionic strength in the surrounding solutions raised by varying ionic salts are seldom looked into. This project focuses on the change of ironic strength and how it affects the CaCO$_3$ crystal polymorphism. There are three common polymorphs of CaCO$_3$, calcite, aragonite, and vaterite. Vaterite is the least stable polymorph of CaCO$_3$ among the three. Its crystal system is hexagonal and possesses a special 3D structure (Hu et al., 2012; Wang and Becker, 2009; Yongchang et al., 2011; Yongchang et al., 2012). The results of this study demonstrate that the ionic strength actually plays a considerable role in polymorph selection; that the percentage of vaterite, among the CaCO$_3$ precipitates is correlated with the concentration of ionic strength of the salt solution.

## METHODOLOGIES

*1. Preparation of salt solution to cultivate CaCO$_3$*

Equal amounts of calcium chloride (CaCl$_2$) and Sodium carbonate (NaHCO$_3$) solutions (4mM) were created and mixed, producing a growth solution containing 2mM Ca$^{2+}$ -CO$_3^{2-}$. The original pH of this solution was decreased by dilute hydrogen chloride to 6.5 and then increased to 9.3 by 1mol/L NaOH to 9.3-9.5 inside a plastic Petri dish, to induce supersaturation of calcium carbonate polymorphs. The pH was increased according to a same trend shown in Fig. 1 to eliminate kinetics differences caused by increase of the pH.

*2. Design of parallel runs to analyze functions of varying ironic additives*

A control run and a series of parallel runs (named "ionic strength" runs) were conducted to analyze different functions of ironic strength of specific ions. For the "ionic strength" runs, in Ca-CO$_3$ salt solutions, various additives (e.g. $NH_4^+$, $Na^+$, $K^+$) of different concentrations were added. Adding both NaCl and Na$_2$SO$_4$ is to compare the effects between divalent and monovalent background electrolytes. Thereafter, Petri dishes were covered by Parafilm, allowed to sit 6-8 hours for calcium carbonate crystals (CaCO$_3$) to grow. Details are described as below:

Control run:
**1.** Prepare 0.002 mol/L Ca-CO$_3$ growth solution using the reagents of CaCl$_2$ and NaHCO$_3$
**2.** Decrease the pH of the solution using 10% HCl to about 6.5
**3.** Increase pH according to the curves shown in Fig. 1 using 1mol/L NaOH to 9.3-9.5

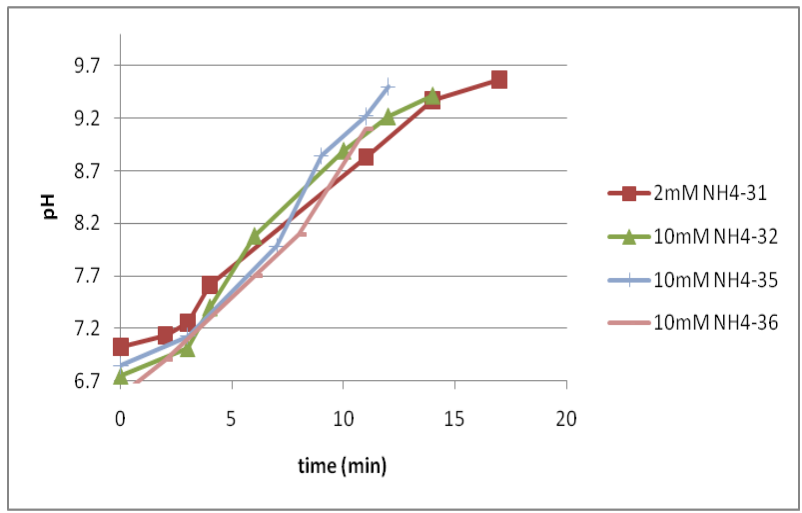

Fig. 1. pH curves

*Ionic strength series 1:*
Preparation of salt solution is the same as control run, except with the addition of 0.0005, 0.005, and 0.01M 1). NaCl, 2). KCl, 3). Na$_2$SO$_4$, and 4).NH$_4$HCO$_3$ respectively in the salt solutions. Every condition is repeated three times, with a total of 12 runs.

*Ionic strength series 2:*
Preparation of salt solution is the same as control run, except adding 0.1 M NaCl, and 1).0.0005, 2). 0.005, and 3). 0.01 M NH$_4$HCO$_3$ respectively in the salt solutions. In total 3 runs.

*3. Analysis of cultivated crystals*

As the crystals are growing, they are also being monitored using an optical microscope, which is a common approach to analyze characteristics of optical materials(Haynie et al., 2009). Once the crystals are fully grown, the top solution in the Petri dishes was poured out and the crystals on the bottom of the dish were then photographed. Once a sufficient amount of CaCO3 crystals was collected, they are analyzed for their crystal structure using x-ray diffraction after being ground into fine pieces using a mortar and pestle, to better understand the crystal polymorphism of the crystals.

The goal of this approach is to cultivate $CaCO_3$ crystals under different chemical but equal kinetic conditions. Subsequently, it is analyzed how various additives influence the polymorph composition of the precipitates.

## RESULTS AND DISCUSSION

Most of the samples from both series follow the expected results of having more vaterite as the ionic concentration increases in the solutions, especially the runs with $NH_4^+$ additives based on the XRD results (Fig. 2). Two experimental cases with ammonium additives followed the correct trend of increasing ionic concentration with increasing vaterite present. The third case had .0005 and .005 M following the correct trend but then 0.01 M showed a dramatic decrease in vaterite composition. However, some runs do not follow this and there are some notable oddities within some of the sample crystals. For example, in the solutions with disodium sulfate ($Na_2SO_4$) in Series One, the vaterite composition decreased as the ionic concentration increased. The two sodium ions seem to have the opposite trend or it could be possibly that the anion is what is affecting this case, as sulfate is not used in any other of the samples.

The 12 cases that had the same amounts of ammonium chloride and disodium sulfate added but with the addition of 0.1 M sodium chloride (NaCl) showed reduced amounts of vaterite than when no sodium chloride was added. This result might demonstrate that additional ions to a solution may reduce the vaterite stability, or that the sodium itself does not stabilize vaterite as well as ammonium or two sodium atoms. The former is suggested because in the previous disodium samples as the ionic concentration increased less vaterite formed. This suggests that it is possible to add "too much" ions to a solution and that a potential "sweet spot" of ionic concentration exists.

## CONCLUSIONS

This study reveals that ionic strength in the growth solutions of $CaCO_3$ is able to influence the polymorphism of $CaCO_3$ precipitation. The major experimental results follow a trend that more vaterite is found among the precipitation as the ionic concentration increases in the solutions; especially the runs added $NH_4^+$. However, for the runs with divalent additives $Na_2SO_4$, there is no clear linear relation of the concentration of additives with the percentage of vaterite, a less stable polymorph of $CaCO_3$.

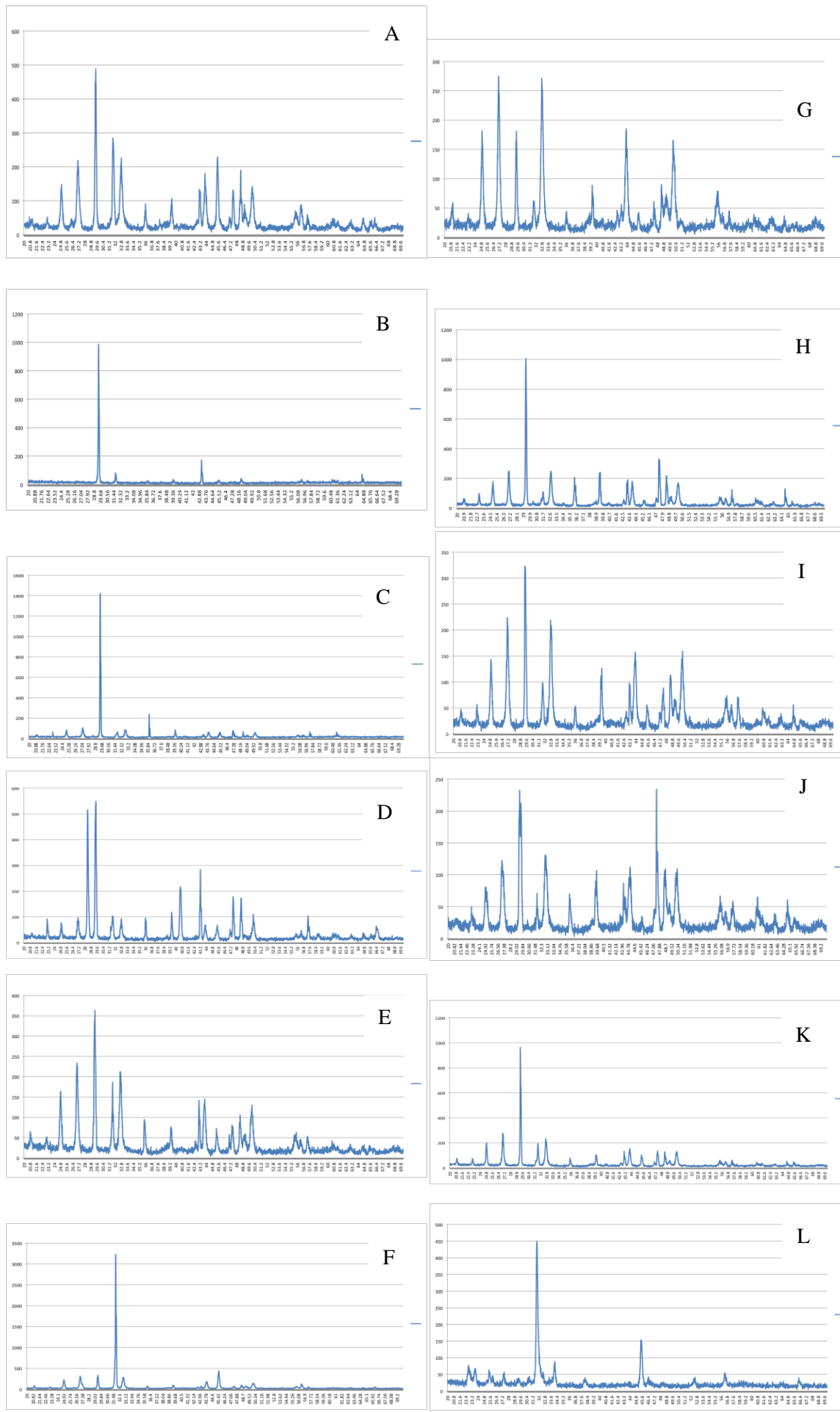

Fig. 2. XRD results of the precipitates under varying conditions of additives. A: NH4 0.01 M, B: Na 0.0005 M, C: K 0.0005 M, D: K 0.01 M, E: NH$_4$ 0.0005 M, F: Na$_2$ 0.0005 M, G: NH$_4$ 0.005 M, H: NH$_4$ 0.01 M, I: NH4 0.005 M, J: NH4 0.01 M, K: Na$_2$ 0.0005 M, L: Na$_2$ 0.005 M